\documentclass[twocolumn,preprintnumbers,a4paper,
superscriptaddress,floatfix,twoside,prd]{revtex4}

\usepackage{bm}
\usepackage{epsfig}
\usepackage{graphics}
\usepackage{natbib}
\usepackage{fancyh}
\usepackage[l]{floatflt}
\usepackage{epsfig}
\usepackage{amssymb}
\usepackage{latexsym}
\usepackage{times}
\usepackage{slashed}
\usepackage{upgreek}
\usepackage{amsmath}
\usepackage{amsfonts}
\usepackage{amsbsy}
\usepackage{amscd}
\usepackage{bbm}

\newcommand{\Eref}[1]{Eq.~(\ref{#1})}
\newcommand{\Erefs}[2]{Eqs.~(\ref{#1}) -- (\ref{#2})}
\newcommand{\Sref}[1]{Sec.~\ref{#1}}
\newcommand{\Fref}[1]{Fig.~\ref{#1}}

\newcommand{\cref}[1]{Ref.~\cite{#1}}
\newcommand{\crefs}[1]{Refs.~\cite{#1}}

\newcommand{\hepth}[1]{{\ftn \tt hep-th/#1}}
\newcommand{\hepph}[1]{{\ftn\tt hep-ph/#1}}

\newcommand{\arxiv}[1]{{\ftn\tt  arXiv:#1}}

\newcommand{\bal}{\begin{align}}
\newcommand{\eall}{\end{align}}
\newcommand{\beqs}{\begin{subequations}}
\newcommand{\eeqs}{\end{subequations}}
\newcommand{\eec}{\end{center}}
\newcommand{\bec}{\begin{center}}

\newcommand{\eem}{\end{matrix}}
\newcommand{\bem}{\begin{matrix}}
\newcommand{\eeq}{\end{equation}}
\newcommand{\beq}{\begin{equation}}
\newcommand{\ba}{\begin{array}}
\newcommand{\ea}{\end{array}}
\newcommand{\bea}{\begin{eqnarray}}
\newcommand{\eea}{\end{eqnarray}}
\newcommand{\baq}{\begin{eqnarray}}
\newcommand{\eaq}{\end{eqnarray}}
\newcommand{\bel}{\begin{align}}
\def\eel{\end{align}}

\newcommand\eqs[2]{Eqs.~(\ref{#1}) and (\ref{#2})}

\newcommand\eqss[3]{Eqs.~(\ref{#1}), (\ref{#2}) and (\ref{#3})}

\newcommand{\ftn}{\footnotesize}

\newcommand{\TeV}{{\mbox{\rm TeV}}}

\newcommand{\GeV}{{\mbox{\rm GeV}}}

\newcommand{\sEref}[2]{Eq.~(\ref{#1}{\ftn\sf {#2}})}

\newcommand{\etal}{{\it et al.\/}}

\def\to{\rightarrow}

\def\lf{\left(}
\def\rg{\right)}
\newcommand\vev[1]{\left\langle {#1} \right\rangle}
\newcommand\veva[1]{\langle {#1} \rangle}
\newcommand{\Gr}{\ensuremath{\widetilde{G}}}

\newcommand{\Vhi}{\ensuremath{V_{\rm H}}}
\newcommand{\tVhi}{\ensuremath{\wtilde V_{\rm H0}}}

\newcommand{\khi}{\ensuremath{K_{\rm H}}}
\newcommand{\khio}{\ensuremath{K_{\rm H0}}}
\newcommand{\whi}{\ensuremath{W_{\rm H}}}
\newcommand{\wkhi}{\ensuremath{\wtilde K}}

\newcommand{\Vhio}{\ensuremath{V_{\rm H0}}}

\newcommand{\mP}{\ensuremath{m_{\rm P}}}

\newcommand{\lm}{\ensuremath{\lambda_\mu}}

\def\openone{\leavevmode\hbox{\small1\kern-3.8pt\normalsize1}}

\newcommand{\dK}{\ensuremath{\Delta K_\mu}}
\newcommand{\km}{\ensuremath{C_H}}

\newcommand{\hd}{{\ensuremath{H_d}}}
\newcommand{\hu}{{\ensuremath{H_u}}}

\newcommand{\zm}{\ensuremath{Z_{-}}}
\newcommand{\tm}{\ensuremath{T_{-}}}

\newcommand{\what}{\ensuremath{\widehat}}
\newcommand{\wtilde}{\ensuremath{\widetilde}}

\def\bz{{\bar Z}}
\def\al{{\alpha}}
\def\Aal{{\bar A}}
\def\Al{{A}}
\def\Bbet{{\bar B}}
\def\Bt{{B}}
\def\bt{{\beta}}

\def\n{\bar{n}}

\def\th{{\theta}}

\newcommand{\mgr}{\ensuremath{m_{3/2}}}
\newcommand{\mgro}{\ensuremath{m_{3/2}}}
\newcommand{\mz}{\ensuremath{\widehat{m}_z}}
\newcommand{\mth}{\ensuremath{\widehat{m}_{\theta}}}
\newcommand{\mzo}{\ensuremath{\widehat{m}_{z}}}

\newcommand{\phc}{\ensuremath{\Phi}}

\def\Ka{K\"{a}hler potential}
\def\Km{K\"{a}hler manifold}
\def\Kaa{K\"{a}hler~}

\newcommand{\Tr}{\mbox{\sf Tr}}

\renewcommand{\refname}{{\bf\scshape References}}

\renewcommand{\thesubsection}{{\small\sf\Alph{subsection}}}

\renewenvironment{subequations}{%
\refstepcounter{equation}%
\setcounter{parentequation}{\value{equation}}%
  \setcounter{equation}{0}
  \def\theequation{\theparentequation{\sf\small\alph{equation}}}%
  \ignorespaces
}{%
  \setcounter{equation}{\value{parentequation}}%
  \ignorespacesafterend
}

\begin{document}

%\preprint{UT-STPD-2/10}

\title{\bf\scshape  Gravity-Mediated SUSY Breaking, \\ {\boldmath $R$} Symmetry and Hyperbolic \Kaa Geometry}

\author{\scshape Constantinos Pallis\\ {\it  Department of Physics and Astronomy, King Saud
University, Riyadh 11451, P.O. Box 2455, SAUDI ARABIA}\\  %{\sl
%e-mail address: }{\ftn\tt kpallis@gen.auth.gr}
}

%School of Technology, Aristotle University of Thessaloniki, GR-541
%24 Thessaloniki, GREECE\\

\begin{abstract}

\noindent {\ftn \bf\scshape Abstract:} A novel realization of the
gravity-mediated SUSY breaking is presented taking into account a
continuous global $R$ symmetry. Consistently with it, we employ a
linear superpotential for the hidden sector superfield and a \Ka\
parameterizing the $SU(1,1)/U(1)$ \Km\ with constant curvature
$-1/2$. The classical vacuum energy vanishes without unnatural
fine tuning and non-vanishing soft SUSY-breaking parameters, of
the order of the gravitino mass, arise. A solution to the $\mu$
problem of MSSM may be also achieved by conveniently applying the
Giudice-Masiero mechanism. The potentially troublesome $R$ axion
may acquire acceptably large mass by explicitly breaking the $R$
symmetry in the \Ka\ through a quartic term which does not affect,
though, the achievements above.
\\ \\ {\scriptsize {\sf PACs numbers: 12.60.Jv, 04.65.+e}
%11.30.Er, 11.30.Pb,
\hfill {\sl\bfseries Published in} {\sl Phys. Rev. D} {\bf 100},
no.~5, 055013 (2019) }
%\pacs{98.80.Cq, 11.30.Qc, 12.60.Jv} it does not require fine tuned parameters

\end{abstract}\pagestyle{fancyplain}

\maketitle

\rhead[\fancyplain{}{ \bf \thepage}]{\fancyplain{}{\sl
Gravity-Mediated SUSY Breaking, $R$ Symmetry and Hyperbolic \Kaa
Geometry}} \lhead[\fancyplain{}{\sl C. Pallis}]{\fancyplain{}{\bf
\thepage}} \cfoot{}

\section{Introduction}\label{intro}

Although still undiscovered, \emph{Supersymmetry} ({\ftn\sf SUSY})
remains one of the most plausible, well-motivated and natural
candidates for the evolution of particle physics beyond the
\emph{Standard Model} ({\ftn\sf SM}). One of the most elusive
problems of the SUSY theories is the mechanism of the SUSY
breaking. According to an elegant and extensively adopted
paradigm, SUSY is spontaneously broken by the \emph{vacuum
expectation values} ({\ftn\sf v.e.vs}) of a set of chiral fields
which form a ``hidden sector'' \cite{nilles} connecting with the
observable sector mostly through gravitational-strength
interactions, including the effects of \emph{Supergravity}
({\ftn\sf SUGRA}).

One of the key ingredients for the successful implementation of
this scenario is the determination of a realistic vacuum for the
relevant SUGRA potential with naturally vanishing or, at least,
tunably small cosmological constant. In view of the recent
scepticism \cite{vafa, lindev} related to the consistency of the
de Sitter vacua within string theory, we here concentrate on the
former possibility proposing a novel gravity-mediated
SUSY-breaking scenario with natural Minkowski solutions at the
classical level. Actually, we improve the well-known Polonyi model
\cite{polonyi} in two directions: Following \cref{hall}, we keep
only the first term of the relevant superpotential which includes
a linear term of the hidden sector field and may become consistent
with a global $R$ symmetry \cite{rnelson} forbidding other terms.
The vanishing of cosmological constant is elegantly addressed by
selecting an appropriate internal space which exhibits a
$SU(1,1)/U(1)$ symmetry \cite{linde1, linde, su11} with constant
curvature $-1/2$. Using a convenient parametrization of the \Km,
which violates though the $R$ symmetry, we show that our model
exhibits novel Minkowski solutions in the context of the
generalized no-scale SUGRA \cite{noscale,noscale1, noscale18}.
Contrary to that case \cite{noscale13}, the gravitino, $\Gr$, mass
is clearly determined at the tree level and the \emph{soft
SUSY-breaking} ({\sf\ftn SSB}) parameters \cite{soft} can readily
acquire adjustable, non-zero values of the order of $\Gr$ mass. We
exemplify these effects, linking the hidden sector to a generic
SUSY model and the \emph{Minimal Supersymmetric SM} ({\ftn\sf
MSSM}). In the latter case, our scheme also offers an explanation
of the $\mu$ term of MSSM by conveniently adapting the
Giudice-Masiero mechanism \cite{masiero}.

% -- cf. \cref{dvali, univ}To some extent,
%Although quite compelling, the above scheme gets into trouble due
%to the presence of the massless $R$ axion.

However, a spontaneously broken continuous and global $R$ symmetry
implies an (pseudo) Nambu-Goldstone boson, the $R$ axion,
\cite{rnelson, raxion} -- as in the case of Peccei-Quinn symmetry
\cite{pq} -- which is cosmologically dangerous. To avoid this
effect, we introduce a quartic term, inspired by \cref{noscale18},
in the \Ka\ which violates $R$ symmetry and allows for
non-vanishing $R$-axion masses without disturbing, though, either
the minimization of the SUGRA potential or the values of the SSB
parameters.

Below, in \Sref{sugra}, we outline  the SUGRA formalism and then
we focus first, in \Sref{hd}, on the hidden sector and then, in
\Sref{obs}, on the visible sector of our model.  Our conclusions
and several perspectives are discussed in \Sref{con}. Possible
connection of our model with no-scale SUGRA is examined in
Appendix A. Unless otherwise stated, we use units where the
reduced Planck mass $\mP=2.433\cdot 10^{18}~\GeV$ is taken to be
unity and charge conjugation is denoted by a bar.

\section{SUGRA Formalism}\label{sugra}

In constructing a SUSY-breaking model based on SUGRA, we mostly
consider two sectors; a so-called hidden sector responsible for
the spontaneous SUSY breaking, and an observable sector which
includes ordinary matter and Higgs fields and which would have
unbroken global SUSY in the absence of the coupling to SUGRA. In
particular, it is assumed that the superpotential has the form
\beq \label{Who} W=W_{\rm H}(Z) + W_{\rm O}\lf\phc_\al\rg,\eeq
in which $W_{\rm H}$ and $W_{\rm O}$ depend only on the chiral
fields of the hidden and observable sectors, respectively. The
hidden sector here consists of just one gauge-singlet superfield
$Z$, similarly to the Polonyi \cite{polonyi} model, whereas the
superfields of the observable sector are denoted by $\phc_\al$.
The suggested \Ka\ may take collectively the form
\beq \label{Kho} K=K_{\rm H}(Z)+\wtilde K(Z)|\phc_\al|^2.\eeq
The specific expressions for $W_{\rm H}$ and $K_{\rm H}$ are given
in \Sref{hd} whereas those for $W_{\rm O}$ and $\wkhi$ in
\Sref{obs}.

Central role in the SUGRA formalism plays the K\"ahler-invariant
function expressed in terms of $K$ and $W$ as follows
\beq \label{Gdef} G = K + \ln |W|^2. \eeq Using it we can derive
the SUGRA scalar potential \beq \label{Vsugra} V = e^{G}\lf
G^{\Al\Bbet} G_\Al G_\Bbet-3\rg=\lf G_{\Al\Bbet}  F^\Al \bar
F^\Bbet-3e^{G}\rg,\eeq where the subscripts denote differentiation
\emph{with respect to} ({\sf\ftn w.r.t}) the fields $Z$ and
$\phc^\al$ and $G^{\Al\Bbet}=K^{\Al\Bbet}$ is the inverse of the
\Kaa\ metric $K_{\Al\Bbet}$. The F terms are defined as
\cite{soft}
\beq \label{Fz} F^\Al = e^{G/2}K^{\Al\Bbet}G_\Bbet
~~~\mbox{and}~~~\bar F^\Aal = e^{G/2}K^{\Aal\Bt}G_\Bt\,.\eeq

The spontaneous SUSY breaking is signaled by the absorption of a
massless fermion named goldstino by $\Gr$, according to the
``super-Higgs'' mechanism, and is accompanied by a non-vanishing
$\Gr$ mass evaluated at the minimum of $V$ as follows \beq
\label{mgr} \mgr=\vev{e^{G/2}}=\frac13\vev{G_{Z\bz}F^Z\bar
F^\bz-V},\eeq where we made use of \Eref{Vsugra} and assume that
$\veva{\phc_\al}\ll\vev{Z}$. The present vacuum energy density
corresponds to $\vev{V}\simeq10^{-120}$, a negligible value w.r.t
the SUSY-breaking mass scale $\mgr
>10^{-15}$ which implies $\veva{F^Z\bar F^\bz} > 10^{-30}$. The extraordinarily
precise cancellation required in \Eref{Vsugra} for fulfilling
simultaneously the two above constraints is the notorious
cosmological constant problem. Since the explanation of the
smallness of $\vev{V}$ is the crucial point of this problem and
the compatibility of the de Sitter solutions with the string
theory is currently under debate \cite{vafa, lindev}, we below
focus on $\vev{V} = 0$ which defines a Minkowski vacuum.
%\vev{e^{\khi/2}\left|W_{\rm H}\right|}=

Under the assumption above, the mass-squared matrices $M_J^2$ of
the particles with spin $J$ composing the final spectrum of the
hidden sector obey the super-trace formula \cite{nilles}
\bea\nonumber {\sf STr}M^2&=&\sum_{J=0}^{3/2}(-1)^{2J} (2J+1)\Tr
M_J^2\\&=&2\mgr^2\veva{G_{Z\bz}^{-2}G_ZG_{\bz}R_{Z\bz}},\label{strace}\eea
where we take into account that $\khi=\khi(Z)$. Also, we define
the Ricci curvature \cite{noscale1,linde1} of the \Km\ as
\beq \label{ricci} R_{Z\bz}=-\partial_Z\partial_\bz\ln {\rm
g}~~~\mbox{with}~~~{\rm g}=\partial_Z\partial_\bz K_{\rm H}\eeq
being the \Kaa\ metric of the hidden space, whose the scalar
curvature is evaluated from the formula \cite{su11}
\beq \label{RH} {\mathcal R}_{\rm
H}=G^{Z\bz}R_{Z\bz}=\lf{\partial_Z{\rm g}\partial_{\bz}{\rm
g}-{\rm g}\partial_Z\partial_\bz{\rm g}}\rg/{{\rm g}^3}.\eeq
%SUSY Breaking
Taking advantage of \eqs{RH}{Vsugra} with $A=Z$ we easily infer
that \Eref{strace} is translated into
\beq \label{strace1} {\sf STr}M^2=6\mgr^2\vev{{\mathcal R}_{\rm
H}},\eeq
which is significantly simplified w.r.t the initial one. E.g., in
the case of the Polonyi model \cite{polonyi} with canonical \Ka\,
we obtain ${\rm g}=1$ and so ${\sf STr}M^2=0$ \cite{nilles}.

\section{Hidden Sector} \label{hd}

In this Section we first -- see \Sref{hd1} -- specify the hidden
sector of our model and then -- see \Sref{hd2} --  investigate the
SUSY-breaking mechanism conserving $R$ symmetry and employing the
curvature of the \Km\ as free parameter. Perturbing mildly the
resulting geometry, we repeat the study, in \Sref{hd3},
considering a convenient $R$-symmetry violating term in the \Ka.

\subsection{\sc\small\sffamily  Model Set-up}\label{hd1}

Taking into account the deep conceptual connection \cite{rnelson}
between $R$ symmetry and SUSY-breaking, we fix \cite{hall} the
form of $W_{\rm H}$ in \Eref{Who} by imposing an $R$ symmetry
under which $Z$ has the $R$ character of $\whi$. Namely, we select
\beq W_{\rm H} = m Z, \label{whi} \eeq
where $m$ is a positive, free parameter with mass dimensions.
Contrary to the Polonyi model \cite{polonyi} and its variants
\cite{olivegr, muray}, we do not consider any $R$-symmetry
violating constant term.

On the other hand, the form of $\khi$ in \Eref{Kho} adopted here
may not be totally $R$ invariant. In particular, we set
\beq K_{\rm H}
=-N\ln\lf1-\frac{|Z|^2-k\zm^n}{N}\rg~~\mbox{with}~~~
\zm=Z-\bz\label{khi}\eeq
and $|Z|<\sqrt{N}$. Here $N, k$ and $n$ are positive free
parameters. Motivated by several superstring and D-brane models
\cite{Ibanez} we consider the integer values of $N$ as the most
natural. We restrict also ourselves to integer $n$'s. In contrast
to the original Polonyi model \cite{polonyi} and its descendents
\cite{hall, olivegr}, where flat internal spaces are assumed,
$\khi$ parameterizes a curved space, with metric
\beqs\beq{\rm g}=Nw\lf N-|Z|^2+k\zm^n\rg^{-2}\label{ds}\eeq
where \beq
w=N-nk^2\zm^{2(n-1)}-k(n-1)\zm^{n-2}\lf\zm^2+n(|Z|^2-N)\rg\label{ds1}\eeq\eeqs
and $n>2$. The $R$-symmetry violation is expressed via $k$ which
is a tiny parameter employed to endow $R$ axion -- see \Sref{hd3}
-- with mass. Small $k$ values are totally natural, in the 't
Hooft's sense \cite{symm}, since nullifying this parameter the $R$
symmetry becomes exact. In the same limit, the \Km\ is totally
$SU(1,1)/U(1)$ symmetric \cite{linde, noscale1, su11} but the
K\"ahler-invariant function -- see \Eref{Gdef} -- still violates
it, due to the form of $\whi$ in \Eref{whi}.

\subsection{\sc\small\sffamily  Totally $R$-Symmetric Case}\label{hd2}

If we set $k=0$ in \Eref{khi} we obtain the exactly $R$-symmetric
version of our model which exhibits an hyperbolic space, in
Poincar\'e disk coordinates \cite{linde, su11}, with metric
\beq\label{ds0} {\rm g}^{(0)}=\partial_Z\partial_\bz K_{\rm H0}
=\lf1-{|Z|^2}/{N}\rg^{-2}\eeq
and constant curvature estimated by \Eref{RH} with result
\beq\label{RHN} {\cal R}_{\rm
H}^{(0)}=-2/N~~~\mbox{since}~~~R^{(0)}_{Z\bz}=-{2}{\rm
g}^{(0)}/N\,.\eeq Here and hereafter, the superscript $(0)$ and
the subscript $0$ denote quantities corresponding to the totally
$R$-symmetric case. The same geometry can be expressed in the
half-plane coordinates as detailed in Appendix A.

\begin{figure}[!t]%\vspace*{-.25in}
\includegraphics[width=60mm,angle=-90]{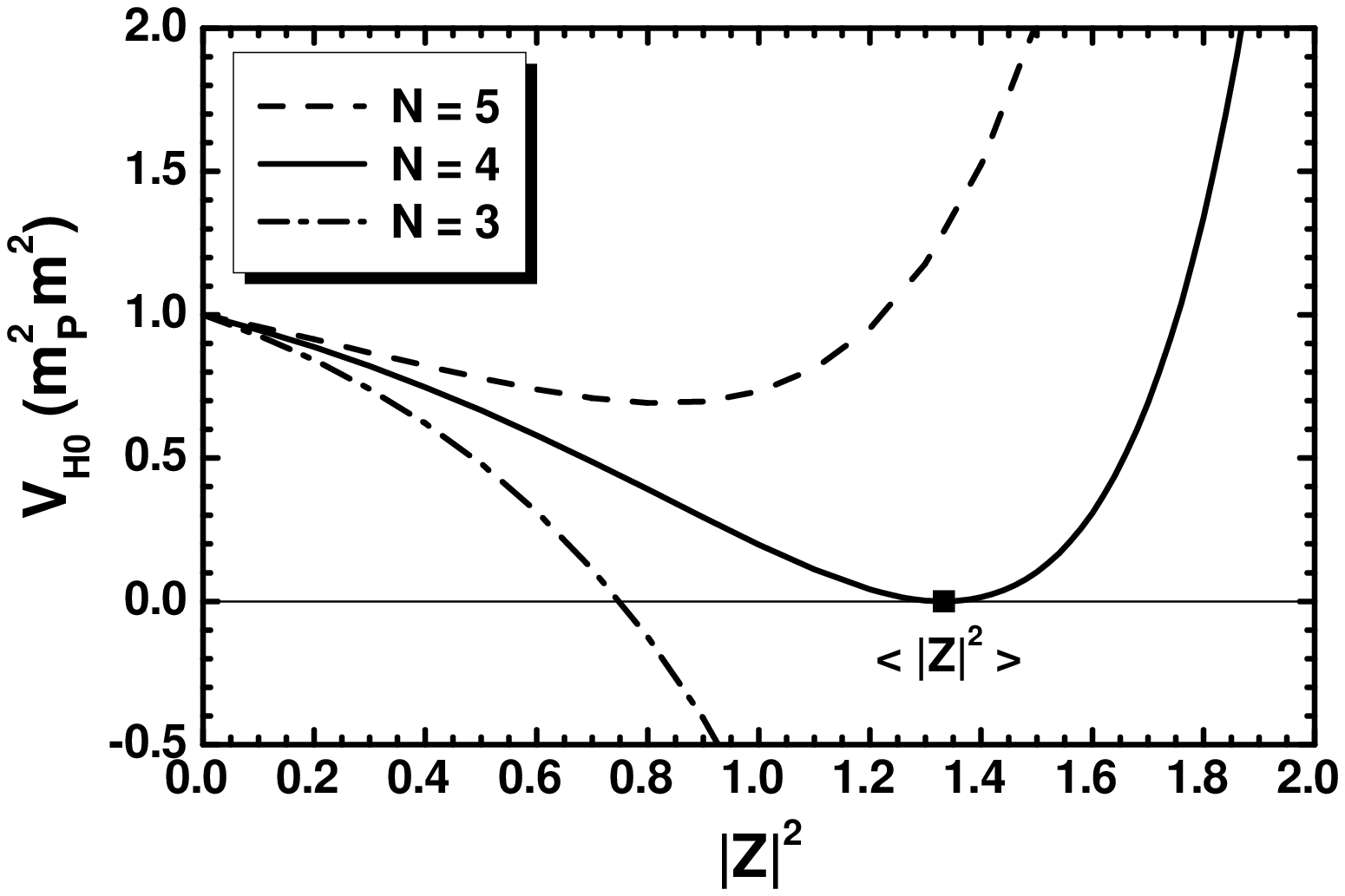}
\caption{\sl \small The (dimensionless) hidden-sector potential
$\Vhio/m^2\mP^2$ in \Eref{Vh} as a function of $|Z|^2$ for $N=3$
(dot-dashed line), $N=4$ (solid line) and $N=5$ (dashed line). The
line $\Vhio=0$ and the value $\vev{|Z|^2}$ are also
indicated.}\label{fig1}
\end{figure}

The corresponding SUGRA potential, $\Vhio$, derived by applying
\Eref{Vsugra}, depends exclusively on $|Z|^2$. Indeed, we obtain
\beq \label{Vh}\Vhio =\lf\frac{m}{N}\rg^2e^{\khio}  \lf\lf N + \lf
N-1\rg |Z|^2\rg^2-3N^2 |Z|^2\rg,\eeq
where we take into account \Eref{ds0} and the equality
\beq \label{Gz}G^{(0)}_Z =\lf\sqrt{{\rm g}^{(0)}}|Z|^2+1\rg/Z=\bar
G^{(0)}_\bz\,. \eeq
We seek below the (preferably integer) value of $N$ that yields a
Minkowski vacuum defined by the conditions
\beq \label{cond} \mbox{\sf\ftn
(a)}~~\vev{\Vhio}=0,~~\mbox{\sf\ftn
(b)}~~\vev{\Vhio'}=0~~~\mbox{and}~~~\mbox{\sf\ftn
(c)}~~\vev{\Vhio''}>0,\eeq
where the derivatives w.r.t $|Z|^2$ are denoted by a prime.
Computing the first derivative of $\Vhio$ in \Eref{Vh} w.r.t
$|Z|^2$, we find
\beq  \label{Vhp} \Vhio'=m^2
\frac{N+(N-1)|Z|^2}{e^{1-\khio}N^3}\Big((N-1)(N-2)|Z|^2-2N\Big).\eeq
%\Big(2 N^2 - N (4 - 5 N + N^2) |Z|^2\\  &&- (N-2) (N-1)^2
%|Z|^4\Big),\eea
Taking into account that $|Z|^2>0$, we infer that \sEref{cond}{b}
implies (for $N\neq1$ and $N\neq2$)
\beq
\label{vevVhp}\vev{\Vhio'}=0~\Rightarrow~\vev{|Z|^2}=\frac{2N}{(N-2)(N-1)},\eeq
which fulfils \sEref{cond}{b} since \beq \label{vevVhpp}
\vev{\Vhio''}=m^2\lf\frac{N-2}{N-3}\rg^{N+1}\lf1-\frac1N\rg^{N+2}>0\eeq
for $N>3$. The value of $\Vhio$ at the minimum is \beq
\label{vevVh}\vev{\Vhio}=m^2\frac{N-4}{N-2}\lf\frac{(N-2)
(N-1)}{(N-3) N}\rg^{N-1}\eeq and can become consistent with
\sEref{cond}{a} for $N=4$. In other words, the value $N=4$ renders
the expression in the parenthesis of \Eref{Vh} equal to the
expansion of a perfect square -- see \Eref{Vhz} below. In view of
\Eref{RHN}, we deduce that the emergence of the Minkowski vacuum
is closely connected with the curvature of the internal space
which is confined to $R^{(0)}_{\rm H}=-1/2$. The structure of
$\Vhio$ in \Eref{Vh} is further highlighted in \Fref{fig1}, where
we depict it for $N=3, 4$ and $5$ (dot-dashed, solid and dashed
line respectively) versus $|Z|^2$. We observe that for \beq
N=4~~~\mbox{and}~~~\vev{|Z|^2}=4/3\label{vev4} \eeq $\Vhio$
exhibits an absolute minimum with vanishing $\veva{\Vhio}$. It is
impressive that this goal is attained without any tuning.
Obviously, tiny, non-zero $\veva{\Vhio}$ can be also achieved by
tuning $N$ to values a little larger than $4$.

%%%%%%%%%%%%%%%%%%%%%%%%%%%%%%%%%%%%%%%%%%%%%%%%%%%%%%%%%%%%%%%%%%%%
%%%%%%%%%%%%%%%%%%%%%%%%%%%%%%%%%%%%%%%%

If we analyze $Z$ according to the description \beq
Z=(z+i\th)/\sqrt{2}\label{Zpara}\eeq and expand $\Vhio$ in
\Eref{Vh} about the configuration
\beq\vev{z}=2
\sqrt{\frac23}~~~\mbox{and}~~\vev{\th}=0,\label{z0}\eeq
-- cf. \Eref{vev4} --, we obtain the hidden-sector spectrum of the
model. This is composed of a massless Nambu-Goldstone boson,
$\th$, -- referred \cite{rnelson} to as an $R$ axion --, a massive
real scalar field, $\what z$, called $R$ saxion, with mass $\mzo$
and the gravitino $\Gr$ -- which absorbs the fermionic partner of
the $R$ saxion, the $R$ axino -- with mass $\mgro$. The former can
be found by substituting \Eref{vevVhpp} with $N=4$ in the formula
\beqs\beq \what m_{z} =\vev{\partial^2_{\what
z}\Vhio}^{1/2}=\vev{\frac{2\Vhio''|Z|^2}{{\rm
g}}}^{1/2}=\frac{3\sqrt{3}}{2}m,\label{mz0}\eeq where we take into
account that $\what z=\sqrt{{\rm g}^{0}}z$ with $\veva{\sqrt{{\rm
g}^{0}}}=\veva{\sqrt{{\rm g}}}=3/2$ -- since along the direction
in \Eref{z0} the $R$-violating term in \Eref{khi} vanishes, we do
not apply the distinction mentioned below \Eref{RHN}. As regards
the $\Gr$ mass, \Eref{mgr} yields \beq \label{mgr0}
\mgro=m\vev{e^{\khi/2}Z}=\frac{3\sqrt{3}}{2}m\,.\eeq\eeqs The
masses above satisfy \Eref{strace1} in view of \eqs{RHN}{vev4},
since
\beq {\sf STr}M_0^2 =\mzo^2-4\mgro^2=-3\mgro^2\,.\label{tr0}\eeq
We see that the mass scale $m$ involved in \Eref{whi} is related
to $\Gr$ mass. Its value is not constrained within our scheme. It
may lie in the range from $\TeV$ until $\mP$ with the former
choice being favored by the resolution of the gauge hierarchy
problem and the latter option being more natural from the point of
view of model building.

Since the $R$ symmetry is explicitly broken by the SSB terms only
in the observable sector, the $R$ axion remains completely
massless if the $R$ symmetry is color, i.e. $SU(3)_{\rm c}$,
\emph{non-anomalous}. To assess the color anomaly we have to know
the complete structure of theory, i.e., the $R$ charges of the
$SU(3)_{\rm c}$ non-singlet fermions -- cf. \crefs{univ,dvali}.
There are model-dependent mechanisms \cite{gaugedR} which may
render the $R$ symmetry anomalous free. In a such case, the
promotion of the global $R$ symmetry to a gauged one surpasses the
difficulty with the massless mode since the $R$ axion is absorbed
by the corresponding gauge boson via the Higgs mechanism. If the
$R$ symmetry is color \emph{anomalous}, non-perturbative QCD
instanton effects \cite{anomalies} result in a mass for the $R$
axion. Since $\vev{z}\sim\mP$, the decay constant of the
$R$-axion, $f_R$, is expected to be of order $\mP$ in
contradiction with the constraint $10^{-8}\lesssim f_R/\mP\lesssim
10^{-6}$ implied by the stellar evolution and the dark matter
abundance in the universe. The constraint on $f_R$ may be
fulfilled, though, considering lower fundamental scale in the \Ka\
-- cf. \cref{ribe}.

In both cases above, another solution to the problem with the
massless $R$ axion is the consideration of $Z$ as a nilpotent
superfield \cite{nilpotent}. In such a case, no sgoldstino
multiplet appears at the SUSY-breaking vacuum and so no $R$ axion
too. Finally, the simplest solution, adopted here, is the explicit
breaking of $R$ symmetry via subdominant terms in $\whi$ and/or
$\khi$ which generates a large enough mass for the $R$ axion. In
particular, its mass must exceed $10~{\rm MeV}$ to evade
astrophysical constraints from production in a supernova
\cite{astro}.

%and at least one more $R$-charged field has to be introduced so as
%to ensure that the $R$ symmetry is broken at high scale and does
%not imply heavy SSB terms for the observable-sector
%superfields

\subsection{\sc\small\sffamily  Including the $R$-Symmetry-Breaking Term}
\label{hd3}

Taking advantage of the nice behavior of $\Vhio$ in \Sref{hd2} we
fix $N=4$ and we allow for non-vanishing $k$ and integer $n$
values in \Eref{khi}. Although no purely theoretical motivation
exists for this term, we can show that $n$ can be uniquely
determined if we require that the resulting SUGRA potential $\Vhi$
takes, along the real direction $\th=0$, the form of $\Vhio$ in
\Eref{Vh} and the $R$ axion becomes massive.

Initially, it is easy to convince ourselves that g in \Eref{ds}
declines from g$^{(0)}$ in \Eref{ds0} for $Z=\bz$ and $n=1$ or
$2$. Therefore, we restrict our analysis to $n\geq3$. Applying
\Eref{Vsugra}, we find that $\Vhi$ takes the form
\beq\Vhi=\frac{m^2}{4}e^{\khi}\Big(
\frac{uv}{w}-12|Z|^2\Big),\label{Vhzz}\eeq where we introduce the
quantities
\beqs\bel \label{u} u&=4+3|Z|^2-k\zm^{n-1}\lf(4n-1)Z+\bz\rg\\
\label{v} v&=4+3|Z|^2+k\zm^{n-1}\lf Z+(4n-1)\bz\rg
\end{align}\eeqs\\ [-0.4cm]
with $u$ and $v$ originating from the numerators of $G_Z$ and
$G_{\bar Z}$ whereas $w$ is given by \Eref{ds1} for $N=4$. Using
the parametrization in \Eref{Zpara}, we can express $\Vhi$ as a
function of $z$ and $\th$ and minimize it in both directions to
determine the Minkowski vacuum. We can show, though, that the
direction $\th=0$ is stable, for $n>3$, and so the Minkowski
vacuum still lies along the direction in \Eref{z0}.

Indeed, $\Vhi$ for $\th=0$ coincides with the one obtained from
\Eref{Vh} for $N=4$, i.e.,
\beq
\Vhi(z,\th=0)=64m^2\frac{(3z^2-8)^2}{(z^2-8)^4}\label{Vhz}\eeq
and therefore, $\vev{z}$ keeps its value in \Eref{z0}. As regards
$\th$, its value in \Eref{z0} satisfies the extremum condition
$\vev{\partial_{\th}\Vhi}=0$ for $n>3$. To prove it, we compute
the first derivative of $\Vhi$ w.r.t $\th$ for $\th=\vev{\th}$
with result
\begin{align}\nonumber \vev{\partial_{\th}\Vhi}&=
-4m^2\lf\frac{3}{2}\rg^4\vev{\frac{\partial_{\th}w}{w}}+\cdots\\
&=
-\frac{27}{2^{2-\frac{n}2}}i^{n-2}n(n-1)(n-2)km^2\vev{\th}^{n-3}+\cdots,\label{vt1}
\end{align} \\ [-4mm]
where the ellipsis represents terms which vanish at the vacuum of
\Eref{z0} for $n>2$. From the expression above, we infer that
\beq \vev{\partial_{\th}\Vhi}=\begin{cases}
-81i\sqrt{2}km^2&\mbox{for}~~n=3;\\
0&\mbox{for}~~n>3.\end{cases}\label{vt2}\eeq
For $n>3$, we can also verify that
\beq \vev{\partial_{z}\partial_{\th}\Vhi}=
\vev{\partial_{\th}\partial_{z}\Vhi}=0. \eeq
if we take into account the following relations
\beqs\bel & \vev{\partial_{z}\partial_{\th}u}=
\vev{\partial_{\th}\partial_{z}u}=\vev{\partial_{z}\partial_{\th}v}=
\vev{\partial_{\th}\partial_{z}v}=0; \\
&
\vev{\partial_{z}\partial_{\th}w}=-(\sqrt{2}i)^{n-2}n(n-1)(n-2)k\vev{\th}^{n-3}\vev{z}\,.\end{align}\eeqs\\
[-4mm]

On the other hand, non-vanishing $R$-axion mass dictates $n=4$.
Indeed, evaluating the second derivative of $\Vhi$ in \Eref{Vhzz}
w.r.t $\th$ for $\th=\vev{\th}$ and taking into account \beq
\vev{u}=\vev{v}=2\vev{w}=8~~\mbox{and}~~\vev{\partial^2_{\th}u}=\vev{\partial^2_{\th}v}=3,\eeq
along with \sEref{cond}{a} which implies
\beq
\vev{uv}=12\vev{w|Z|^2}~~\stackrel{(\ref{z0})}{\Longrightarrow}~~\vev{uv}=16\vev{w},\label{mcond}\eeq
we arrive at the following result
\begin{align}\nonumber \vev{\partial_{\th}^2\Vhi}&=
{m^2}\lf\frac{3}{2}\rg^4\Bigg(4\vev{\frac{\partial^2_{\th}u}{u}+\frac{\partial^2_{\th}v}{v}-\frac{\partial^2_{\th}w}{w}}-3\Bigg)\\
&=
-\frac{27i^{n-2}}{2^{2-\frac{n}{2}}}n(n-1)(n-2)(n-3)km^2\vev{\th}^{n-4}.\label{vt3}
\end{align} \\ [-4mm]
The expression above assumes a positive value for $n=4$, whereas
it vanishes for $n>4$. Canonically normalizing the relevant mode,
we may translate the above output as follows
\beq \mth=\vev{\partial^2_{\what \th}\Vhi}^{\frac12}
=\vev{\partial^2_{\th}\Vhi/{\rm g}}^{\frac12}=\begin{cases}
12\sqrt{2k}m&\mbox{for}~~n=4,\\0&\mbox{for}~~n>4.\end{cases}
\label{mth}\eeq
Consequently, setting $n=4$ and $k>0$ in \Eref{khi} does not
modify $\Vhi$ from $\Vhio$ in the real direction but just allows
for a non-vanishing $R$-axion mass. Note that the same (quartic)
term is also employed in \cref{noscale18} to stabilize the
imaginary direction of the SUSY breaking field within a
no-scale-type model. The strength of the $R$ symmetry breaking is
adequate to render $\what\th$ heavier than a few tens of MeV
freeing it, thereby, from the astrophysical constraints. E.g., for
$m=1~\TeV$, it is enough to take $k\geq4\cdot10^{-13}$ -- where we
restore the units for convenience.

\begin{figure}[t]%
\epsfig{file=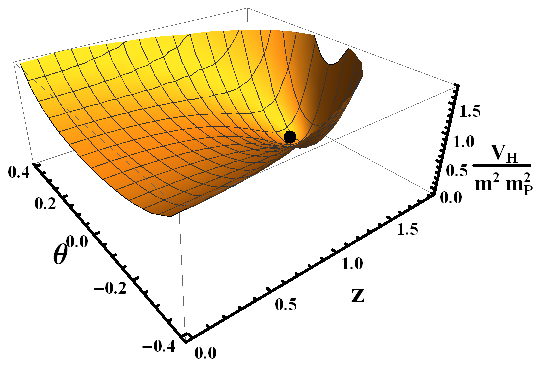,width=8.7cm,,angle=-0} \vspace*{2.3in}
\caption{\sl \small Three dimensional plot of the (dimensionless)
hidden-sector potential $V_{\rm H}/m^2 \mP^2$ in \Eref{Vhzz} for
$n=4$ and $k=0.1$ as a function of the parameters $z$ and $\th$
defined in \Eref{Zpara}. The location of the Minkowski vacuum in
\Eref{z0} is also depicted by a thick black point.}\label{fig5}
\end{figure}

The conclusions of the analysis above can be also verified by
\Fref{fig5}, where we display the relevant three dimensional plot
of the dimensionless quantity $\Vhi/m^2\mP^2$ given by \Eref{Vhzz}
for $k=0.1$ and $n=4$ versus $z$ and $\th$. We see that the
direction $\th=0$ is a valley of minima, along which the
minimization of $\Vhi$ w.r.t $z$ may be safely performed. As a
consequence, the Minkowski vacuum in \Eref{z0} indicated by the
black thick point is also included in this path.

%%%%%%%%%%%%%%%%%%%%%%%%%%%%%%%%%%%%%%%%%%%%%%%%%%%%%%%%%%%%%%%%%%%%
%%%%%%%%%%%%%%%%%%%%%%%%%%%%%%%%%%%%%%%%

Besides the $R$ axion, $\what \th$, which is massive for $n=4$ and
$k>0$ in \Eref{khi}, the particle spectrum of the present version
of our model comprises also $\what z$ and $\Gr$ whose the masses
are given by \eqs{mz0}{mgr0} respectively since the $z$-dependent
form of $\Vhi$ in \Eref{Vhzz} coincides with that of $\Vhio$ in
\Eref{Vh}. We can verify that these masses obey \Eref{strace1}
with ${\mathcal R}_{\rm H}$ estimated by \Eref{RH} with result
\beq\label{RHk}  {\mathcal R}_{\rm
H}=-\frac12\lf1+\frac{3k}{4}\lf|Z|^2-4\rg^3\rg\eeq
%where we take into account that \beq \label{Rzzk}
%R_{Z\bz}=\frac{8(48k-1)-288kz^2+72kz^4-6kz^6}{(z^2-4)^2}\eeq
for $Z=\bz$. Indeed, evaluating $\vev{{\mathcal R}_{\rm H}}$ we
end up with
\beq {\sf STr}M^2
=\mz^2+\mth^2-4\mgr^2=\frac{1}{3}\lf128k-9\rg\mgr^2\,.\label{trkl}\eeq
Checking the hierarchy of the various masses, we infer that \beq
\mz=\mgr~~~\mbox{and}~~~\mth\leq\mgr~~~\mbox{for}~~~k\leq\frac18\sqrt{\frac32}.
\eeq Therefore, no decay of $\what z$ and $\what \th$ (for the
$k$'s above) into $\Gr$ is allowed in contrast to the models with
strongly stabilized sgoldstino -- cf. \crefs{olivegr, olivepheno,
strongpheno}. As a consequence, no extra contribution to the relic
abundance of $\Gr$ before nucleosyntesis arises and no extra
constraint has to be imposed on the reheat temperature --
cf.~\cref{raxion}.

Let us, finally, note that the problem of the vanishing $R$-axion
mass can be also solved, if we set the quartic term in \Eref{khi}
outside argument of the logarithm there. In particular, if we
adopt one of the $K_{\rm H}$ below
\beqs\bel K_{\rm H} &=-4\ln\lf1-|Z|^2/4\rg-N_k\ln\lf1+k\zm^4/N_k\rg,\label{kh1} \\
K_{\rm H} &=-4\ln\lf1-|Z|^2/4\rg-k\zm^4,
\label{kh2}\end{align}\eeqs
the $R$-axion acquires mass \beq \mth= 8\sqrt{3k}m, \eeq which is
similar to that found in \Eref{mth}. The prefactor $N_k$ in
\Eref{kh1} remains an undetermine positive constant.

\section{Observable Sector}\label{obs}

In this section we specify the transmission of the SUSY breaking
to the observable sector of SUSY models. We consider first, in
\Sref{obs1}, a generic SUSY model and then, in \Sref{obs2}, we
focus on the MSSM proposing a solution to the $\mu$ problem. Since
the quantities of the hidden sector related to the present set-up
are computed exclusively at the Minkowski vacuum in \Eref{z0}, the
results are obviously independent from the violation of the $R$
symmetry.

\subsection{\sc\small\sffamily Generic Model}\label{obs1}

To investigate the response of the visible sector to the invisible
one, introduced in \Sref{hd}, we have to specify $W_{\rm O}$ and
$\wkhi$ in \eqs{Who}{Kho}. We here adopt the following, quite
generic form
\beq W_{\rm O}=h \phc_1\phc_2\phc_3+\mu\phc_4\phc_5,
\label{w0}\eeq
where we assign $R$ charge $2/3$ for each of $\phc_1, \phc_2$ and
$\phc_3$ and $1$ for each of $\phc_4$ and $\phc_5$ -- let assume
that $W$ and $Z$ carry $R$ charge $2$.  We also consider that
$\phc_\al$ with $\al=1,...,5$ are involved in one of the following
\Ka s
\beqs\bel
K_1&=\khi+\mbox{$\sum_\al$}|\phc_\al|^2,~~~\label{K1}\\
K_2&=-4\ln\Big(1-\lf|Z|^2-k\zm^4+\mbox{$\sum_\al$}|\phc_\al|^2\rg/4\Big),\label{K2}\\
K_3&=\khi-N_{\rm O}\ln\left(1-\mbox{$\sum_\al$}|\phc_\al|^2/N_{\rm
O}\right),\label{K3} \end{align}\eeqs
where $\khi$ is given by \Eref{khi} for $n=4$ and the specific
value of $N_{\rm O}>0$ is irrelevant for our purposes. We also
restrict ourselves to universal SSB parameters, i.e., the same for
any $\phc_\al$. If we expand the $K$'s above for low $\phc_\al$
values, these may assume the form shown in \Eref{Kho}, with
$\wkhi$ being identified as
\beq \label{wk} \wkhi=\begin{cases} 1&\mbox{for}~~K=K_1,
K_3;\\\lf1-(|Z|^2-k\zm^4)/4\rg^{-1}&\mbox{for}~~K=K_2\,.\end{cases}\eeq

Replacing $Z$ by its v.e.v, \Eref{z0}, in the total SUGRA
potential, \Eref{Vsugra}, and take $\mP\to\infty$ keeping $\mgr$
fixed, we obtain the SSB terms in the effective low energy
potential which can be written as
\beq V_{\rm SSB}= \wtilde m_\al^2 |\what\phc_\al|^2+\lf Ah
\what\phc_1\what\phc_2\what\phc_3+ B\mu
\what\phc_4\what\phc_5+{\rm h.c.}\rg, \label{vsoft} \eeq
where the canonically normalized fields
$\what\phc_\al=\veva{\wtilde K}^{1/2}\phc_\al$ are denoted by hats
and the SSB parameters may be found by adapting the general
formulae of \cref{soft} to our case. I.e.,
\beqs\bel
\wtilde m_\al^2&=\mgr^2-\vev{\bar F^\bz F^Z\partial_\bz\partial_Z
\ln \wtilde K};\label{ma}\\
A&=\vev{e^{\khi/2}\wtilde K^{-3/2}F^Z\lf
\partial_Z\khi-\partial_Z\ln \wtilde K^{3}\rg};~~~\label{Aa} \\
B&=\vev{{e^{\khi/2}/\wtilde K}\bigg(
F^Z\lf\partial_Z\khi-\partial_Z\ln \wtilde
K^{2}\rg-\mgr\bigg)}.\label{Ba}
\end{align}\eeqs \\ [-0.4cm]
Note that $h$ and $\mu$ are considered as independent of $Z$ and
remain unhatted in \Eref{vsoft} -- cf.~\cref{soft}. In deriving
the values of the SSB parameters above, we find it convenient to
distinguish the cases:

\subparagraph{\sl (a)} For $K=K_1$ and $K_3$, we see from
\Eref{wk} that $\wkhi$ is constant and so  the relevant
derivatives are eliminated. Substituting
\beq \vev{F^Z}=\veva{\bar
F^\bz}=\frac{2\mgro}{\sqrt{3}},~\vev{e^{\khi/4}}=\frac32,~\vev{\partial_Z\khi}=\sqrt{3}\label{aux1}\eeq
into \Erefs{ma}{Ba}, we arrive at
%
%\bea
\beq \wtilde
m_\al=\mgro~~~\mbox{and}~~~A=2B=\frac92\mgro\,.\label{mk1}\eeq

\subparagraph{\sl (b)} For $K=K_2$, $\wkhi$ in \Eref{wk} is $Z$
dependent with $\veva{\wkhi}=3/2$ and the relevant derivatives are
found to be
\beq
\vev{\partial_Z\ln\wkhi^2}=\frac{2}{3}\vev{\partial_Z\ln\wkhi^3}=\frac{\sqrt{3}}{2},~
\vev{\partial_\bz\partial_Z\ln\wkhi}=\frac{9}{16}.\label{aux2}\eeq
Inserting the expressions above into \Erefs{ma}{Ba} we end up with
\beq \wtilde m_\al=\frac12\mgro,~A=\frac12\sqrt{\frac32}\mgro
~~~\mbox{and}~~~B=0,\label{mk2}\eeq
%\eea
where the last result arises from a cancellation in the last
factor of \Eref{Ba}.

\subparagraph{} Let us emphasize, finally, that $U(1)_R$ is
totally broken for $k\neq0$ in \Eref{khi} and so, no topological
defects are generated when $Z$ acquires its v.e.v in \Eref{z0}.
For $k=0$ the terms in $V_{\rm SSB}$ explicitly break $U(1)_R$ to
its subgroup $\mathbb{Z}_2^{R}$. Since $Z$ has the $R$ symmetry of
$\whi$, $\vev{z}$ in \Eref{z0} breaks also spontaneously $U(1)_R$
to $\mathbb{Z}_2^{R}$. Thanks to this fact, $\mathbb{Z}_2^{R}$
remains unbroken and so, no disastrous domain walls are formed in
this case too.

\subsection{\sc\small\sffamily MSSM}
\label{obs2}

%The connection of this hidden sector with  causes concerns about
%the ability of an explanation of the $\mu$ term of  MSSM in
%accordance with the imposed $R$ symmetry.

Trying to combine $\whi$ in \Eref{whi} with an even more realistic
observable sector we consider MSSM and we show how the SUSY
breaking is communicated to the scalar and gaugino sector in
Secs.~\ref{mssm1} and \ref{mssm2} respectively.

\subsubsection{\small\sf Scalar Sector -- Generation of the $\mu$ Term} \label{mssm1}

As shown in \eqs{mk1}{mk2}, the existence of the bilinear term in
\Eref{vsoft} is relied on the introduction of the similar term in
$W_{\rm O}$. In the case of MSSM, such a term, involving the Higgs
superfields $H_u$ and $H_d$ coupled to the up and down quark
respectively, with $\mu\sim 1~\TeV$ is crucial for the electroweak
symmetry breaking and the generation of masses for the fermions.
However, we would like to avoid the introduction by hand of a low
energy scale into the superpotential of MSSM, $W_{\rm MSSM}$. To
achieve that, we assign $R$ charges equal to $2$ for both $H_u$
and $H_d$ whereas all the other fields of MSSM -- i.e., $i$th
generation $SU(2)_{\rm L}$ doublet left-handed quark and lepton
superfields, $Q_i$ and $L_i$, and the $SU(2)_{\rm L}$ singlet
antiquark $u^c_i$ and ${d_i}^c$ and antilepton superfields and
$e^c_i$ -- have zero $R$ charges. Note that these $R$ assignments
prohibit not only the term $\mu\hu\hd$ but also a term $\lm
Z\hu\hd$ which leads to unacceptable phenomenology since
$\mu\sim\vev{Z}$. Consequently, the resulting $W_{\rm MSSM}$
exhibits the structure of $W_{\rm O}$ in \Eref{w0} with $\mu=0$,
i.e.,
\bea \nonumber W_{\rm MSSM} &=& h_{D} {d}^c {Q} \hd + h_{U} {u}^c
{Q} \hu+h_{E} {e}^c {L} \hd \\ &=&\frac16 h_{\al\bt\gamma}
\phc_\al\phc_\bt\phc_\gamma\,,\label{wmssm}\eea
where we suppress the generation indices, consider real values of
$W_{\rm MSSM}$ for simplicity and set $h_{\al\bt\gamma}=h_I$ with
$I=D,u,E$. The resulting $R$ symmetry is anomalous since the $R$
color anomaly, defined as the sum of the $R$ charges over the
$SU(3)_{\rm c}$ non-singlet fermions of the theory, is $N_R=12$ --
i.e., $U(1)_R$ is broken by the QCD instanton effects down to its
$\mathbb{Z}_{12}^{R}$ subgroup. As a consequence, the $R$ axion is
cosmologically safe if it becomes adequately massive, i.e., if the
$U(1)_R$ is explicitly violated. Thanks to this violation, no
domain walls are formed too.

Despite the fact that no mixing between $\hu$ and $\hd$ exists in
$W_{\rm MSSM}$, in \Eref{wmssm}, such a term emerges in the part
of the potential including the SSB terms
\bea \nonumber V_{\rm SSB}&=& \wtilde m_\al^2
|\what\phc_\al|^2+\lf\frac16 A_{\al\bt\gamma} h_{\al\bt\gamma}
\what\phc_\al\what\phc_\bt\what\phc_\gamma\right. \\&+& \wtilde
B\mu \what H_u\what H_d+{\rm h.c.}\Big), \label{vmssm} \eea
if we add (somehow) to the $K$'s in \Erefs{K1}{K3} the following
higher order terms, inspired by \cref{masiero},
\beq \dK=\lm\frac{\bz^{2}}{\mP^2}\hu\hd\ +\ {\rm
h.c.},\label{dK}\eeq
where $\lm$ is a real constant and $\what\phc_\al$ in \Eref{vmssm}
are related to the unhatted ones as shown below \Eref{vsoft}.  Due
to the adopted $R$ symmetry, the terms in \Eref{dK} are one order
of magnitude higher than those proposed in the original paper
\cite{masiero}. However, we show below that the magnitude of the
resulting $\wtilde B\mu$ is of the correct order of magnitude.

To be more specific, we consider the following alternative \Ka s
\beqs\bel
K_{11}&=K_1+\dK,~~~\label{K11}\\
K_{21}&=K_2+\dK,\label{K21}\\
K_{22}&=-4\ln\left(1-\left(|Z|^2-k\zm^4+\mbox{$\sum_\al$}|\phc_\al|^2-\dK\right)/4\right),~~~~~\label{K22}\\
K_{23}&=-4\ln\Big(1-\left(|Z|^2-k\zm^4-\dK\right)/4\Big)+\mbox{$\sum_\al$}|\phc_\al|^2,~~~~~~~\label{K23}\end{align}\eeqs
\\ [-0.4cm]
where $K_1$ and $K_2$ are defined in \eqs{K1}{K2} respectively.
The $K$'s above may be brought into the form
\beq K_{\rm MSSM}=K+\big(\km\hu\hd+{\rm h.c.}\big),
\label{kmssm}\eeq
where $K$ is defined in \Eref{Kho}, with
\beqs\beq \label{wkhi1} \wkhi=\begin{cases}
1&\mbox{for}~~K=K_{11}, K_{23};\\\lf1-(|Z|^2-k\zm^4)/4\rg^{-1}
&\mbox{for}~~K=K_{21}, K_{22},\end{cases} \eeq
and $\km$ is found by expanding the $K$'s in \Erefs{K11}{K23} for
low $\hu$ and $\hd$ values with result \beq \label{km}
\km=\lm\frac{\bz^{2}}{\mP^2}\begin{cases} 1&\mbox{for}~~K=K_{11},
K_{21};\\
\Big(\frac{|Z|^2-k\zm^4}{4}-1\Big)^{-1}&\mbox{for}~~K=K_{22},K_{23}.\end{cases}
\eeq \eeqs
Thanks to non-vanishing $\km$, we expect that the effective
coefficient $\wtilde B\mu$ in \Eref{vmssm} assumes a
non-vanishing, in principle, value which may be found by applying
the formula \cite{soft}
\bea\nonumber\wtilde B\mu&=&
\frac{\mgr}{\veva{\wkhi}}\Bigg(2\mgr\vev{\km}-\vev{\bar F^\bz
\partial_\bz\km}+\vev{F^Z
\partial_Z\km}\\ \nonumber&+&
\frac{1}{\mgr}\vev{\bar F^\bz F^Z\partial_\bz\km\partial_Z\ln
\wtilde K^2-\bar F^\bz F^Z\partial_\bz\partial_Z\km}\\ &-&
\vev{F^Z\km\partial_Z\ln \wtilde K^2}\Bigg).\label{mugen}\eea

Making use of \eqs{ma}{Aa} we extract the following SSB parameters
\beqs\beq \label{mA} \frac{\wtilde m_\al}{\mgro}=\begin{cases} 1\\
\frac12
\end{cases}\hspace{-0.2cm}\mbox{and}~~~\frac{A_{\al\bt\gamma}}{\mgro}=
\begin{cases}~~\frac92\\\lf\frac38\rg^{\frac12}
\end{cases}\hspace{-0.2cm}\mbox{for}~~K= \begin{cases} K_{11}, K_{23};\\
K_{21}, K_{22},\end{cases}\eeq
as expected if we compare the $K$'s in \Erefs{K11}{K23} with those
in \Erefs{K1}{K3}. As regards $\wtilde B\mu$, \Eref{mugen} yields
\beq \label{Bm} \frac{\wtilde B\mu}{\mgr^2}= \lm\begin{cases}0 &\mbox{for}~~K=K_{11};\\ 8/9&\mbox{for}~~K=K_{21};\\
2/3&\mbox{for}~~K=K_{22};\\ 4&\mbox{for}~~K=K_{23},
\end{cases}\eeq\eeqs
where we take into account the following:

\subparagraph{\sl (a)} For $K=K_{11}$ and $K_{21}$, $\km$ in
\Eref{km} is only $\bz$ dependent and so we have
\beqs\beq
\vev{\km}=4\lm/3,~\vev{\partial_Z\km}=0~~\mbox{and}~~\vev{\partial_\bz\km}=4\lm/\sqrt{3}.\label{aux3}\eeq
As a consequence, a cancellation occurs in the two first terms of
the right-hand side of \Eref{mugen} for $K=K_{11}$ yielding
$\wtilde B\mu=0$. For $K=K_{21}$, derivatives involving $\wkhi$
can be computed with the aid of \Eref{aux2}.

\subparagraph{\sl (b)} For $K=K_{22}$ and $K_{23}$, $\km$ in
\Eref{km} is both $Z$ and $\bz$ dependent and so we estimate
\bea \vev{\km}&=&2\vev{\partial_\bz\partial_Z\km}/3=-2\lm;\label{aux4} \\
\vev{\partial_Z\km}&=&\vev{\partial_\bz\km}/5=-\sqrt{3}\lm/2.\label{aux5}\eea\eeqs
For $K=K_{22}$, $\wkhi$ is non trivial and its contribution into
\Eref{mugen} is computed by employing \Eref{aux2}.

In conclusion, the $\mu$ term of MSSM can be generated
consistently with the imposed $R$ symmetry for $K=K_{21}, K_{22}$
and $K_{23}$ in \Erefs{K21}{K23}.

%\newpage

\subsubsection{\small\sf Gaugino Sector}\label{mssm2}

Apart from the SSB terms for the scalars, we can also obtain
masses $M_{\rm a}$ for the (canonically normalized) gauginos
$\what \uplambda_{\rm a}$ -- where ${\rm a}=1,2,3$ runs over the
factors of the gauge group of MSSM, $U(1)_Y$, $SU(2)_{\rm L}$ and
$SU(3)_{\rm c}$ with gauge coupling constants $g_{\rm a}$
respectively. These depend not only on $K_{\rm H}$ and $W_{\rm H}$
but also on the selected gauge-kinetic function $f_{\rm a}$ which
is an holomorphic dimensionless function of the chiral
superfields. Adapting to our case the most general formula
\cite{kapnilles},  we find that the ratios of the running gaugino
masses $M_{\rm a}$ over the gauge coupling constants squared
$g_{\rm a}^2$ at a renormalization point are given by
\bea\nonumber\frac{M_{\rm a}}{g_{\rm a}^2}&=& \frac12
\vev{F^Z\partial_Z f_{\rm a}}+\frac{1}{64\pi^2}C_{\rm
a}\partial_Z\ln\mbox{Re}(f_{\rm a})\\&+&\frac1{16\pi^2}b_{\rm
a}\lf\mgr+\frac13\vev{F^Z\partial_Z \khi}\rg\nonumber
\\&-&\frac1{8\pi^2}\sum_a C^a_{\rm
a}\vev{(F^Z)^2\partial_Z \ln\lf e^{-\khi/3}\wkhi\rg}. \label{Ma}
\eea
Here $(b_{\rm a})=(33/5,1,-3)$ are the one-loop beta function
coefficients, $C_{\rm a}$ are the quadratic Casimir of the gauge
multiplets and $(C_{\rm a}^a)=(33/5,7,6)$ are the quadratic
Casimir of the representations $\Phi^a$ of MSSM.

Since the gauginos carry R-charge $+1$, the Majorana gaugino mass
terms originating from a polyonymic form of $f_{\rm a}$ violate
strongly the $R$ symmetry in the SUGRA Lagrangian --
cf.~\cref{gauginoHall} -- and generate potentially dangerous
radiative corrections \cite{corr} to $\khi$ in \Eref{khi}. For
this reason, we may assume that $f_{\rm a}$ is a constant.
However, the remaining contributions to $M_{\rm a}$ from gauge
anomalies are loop suppressed and violate mildly the $R$ symmetry.
For $m\ll\mP$ and all possible $K$'s in \Erefs{K11}{K23} we obtain
\beq \label{Maan} \frac{M_{\rm a}}{\mgr}= \frac1{\pi^2}\begin{cases}11/16 &\mbox{for}~~{\rm a}=1;\\
5/48&\mbox{for}~~{\rm a}=2;\\-5/16&\mbox{for}~~{\rm a}=3\,,
\end{cases}\eeq
where we make use of \Eref{aux1}. Note that the last term in
\Eref{Ma} turns out to be suppressed by $\mgr^2/\mP$. As in the
original anomaly mediated scenario, the gaugino corresponding to
${\rm a}=2$ tends to be the lightest one and the $M_{\rm a}$'s
turn out to be one order of magnitude lower than $\mgr$ or
$\widetilde m_\alpha$ due to the large denominators.

%We ignore here
%contributions to $M_{\rm a}$ via gauge anomalies which are treated
%as an alternative source of non-vanishing $M_{\rm a}$.

\section{Conclusions and Perspectives} \label{con}

We presented an improved version of the well-known Polonyi model
using as guideline a global $R$ symmetry which is badly violated
in the superpotential of that model. As a starting point, we
investigated a theory completely consistent with this $R$ symmetry
-- which uniquely determines the superpotential in \Eref{whi} --
selecting a specific hyperbolic geometry for the \Km\ with metric
given by \Eref{ds0}. Constraining the curvature of this space to a
natural value -- see \Eref{vev4} --, from the point of view of the
string theory, we succeeded to minimize the relevant SUGRA
potential at a SUSY-breaking Minkowski vacuum. The presence of the
cosmologically dangerous $R$ axion in the spectrum of the model
can be eluded by including a quartic term in the \Ka\ -- i.e.,
setting $n=4$ in \Eref{khi} -- which breaks the $R$ symmetry
without modifying the SUGRA potential, along its real direction,
and the position of the Minkowski vacuum in \Eref{z0}. No
string-theoretical origin can be invoked for this term, though.

It is gratifying that the $R$ saxion and axion may acquire masses
lower than  or equal to the $\Gr$ mass and so the $\Gr$ problem is
not aggravated. The model communicates the SUSY breaking to the
visible world, allowing for non-vanishing SSB (i.e. soft
SUSY-breaking) parameters which do not depend on the $R$-violating
term. More specifically, the SSB masses for the scalars are of the
order of $\mgr$ whereas those for gauginos may be one order of
magnitude lower, originating from gauge anomalies. Furthermore,
the consideration of a higher order non-holomorphic term in the
\Ka\ -- see \Eref{dK} -- offers an explanation of the $\mu$
problem of MSSM inspired by the Giudice-Masiero mechanism.

In its current realization, our model does not support viable
inflation driven by $Z$, mainly due to low scalar spectral index
achieved in small-field inflationary models. However, it can be
combined with an inflationary sector compatible with the $R$
symmetry -- see, e.g., \crefs{dvali, univ}. In a such situation we
expect that $Z$ is displaced from its v.e.v in \Eref{z0} to lower
values due to the large mass that it acquires during inflation and
rolls towards its v.e.v after it -- see, e.g.,
\crefs{buch,olivepheno,eev,kingpolonyi,ketov,riotto}. In the
course of the decaying-inflaton period which follows inflation,
$Z$ tracks an instantaneous minimum \cite{adiabatic} until the
Hubble parameter becomes of the order of its mass. Successively it
starts to oscillate about its v.e.v. in \Eref{z0} and may or may
not dominate the Universe, depending on the initial amplitude of
the coherent oscillations. The latter possibility is more favored,
since it does not dilute any preexisting lepton asymmetry and does
not disturb the success of the Big Bang nucleosynthesis
\cite{polonyiproblem}. It can be facilitated if $R$ saxion is
strongly stabilized through a large enough higher order term of
the \Ka\ \cite{olivegr}, or if it participates there in a strong
enough coupling with the inflaton \cite{adiabatic}. Obviously such
complications may affect our scheme and deserve further
investigation. Moreover, the $R$ axion is expected to be stable on
cosmological time scales due to weak decay widths \cite{ribe}. It
would be premature, though, to say anything about its candidacy as
dark matter particle before clarify the fate of the $R$ saxion.

Another prospect of our setting is related to the low-energy SUSY
searches. Indeed, the values for SSB parameters found in
\Sref{obs2} may be used as boundary conditions imposed at a high
scale in order to solve the renormalization group equations which
govern their evolution up-to a low scale. Finding their values
there, we can impose radiative electroweak symmetry breaking,
derive the sparticle spectrum  and check its compatibility with a
number of phenomenological requirements -- cf.~\crefs{noscale13,
strongpheno, queve}. The viability of our scheme against these
constraints is an important open issue. The fact that the majority
of the SSB parameters gain values of the same order of magnitude
helps to this direction. Possible non universalities, caused by
associating different $\wkhi$'s to $\phc_\al$ may further
facilitate the achievement of acceptable results.

Despite the uncertainties above, we believe that the introduction
of a novel model for SUSY breaking without tuning can be
considered as an important development which offers the
opportunity for further explorations towards several
cosmo-phenomenological directions.

%\newpage

\paragraph*{\small\bfseries\scshape Acknowledgments} {\small I would like to acknowledge
F.~Fa-rakos, G. Lazarides and Q. Shafi for useful discussions.
This project was supported by King Saud University, Deanship of
Scientific Research, College of Sciences Research Center.}

\appendix

\renewenvironment{subequations}{%
\refstepcounter{equation}%
% \theparentequation{\theequation}%
\setcounter{parentequation}{\value{equation}}%
  \setcounter{equation}{0}
  \def\theequation{A\theparentequation{\small\sffamily\alph{equation}}}%
  \ignorespaces
}{%
  \setcounter{equation}{\value{parentequation}}%
  \ignorespacesafterend
}
\renewcommand{\thesubsection}{{\small\sf\arabic{subsection}}}

\section{Half-Plane Parametrization of Hyperbolic Geometry}

In this Appendix we employ an alternative parameterization of
hyperbolic geometry which, although violates the $R$ symmetry,
allows us to compare our model with similar ones established in
the context of generalized no-scale SUGRA
\cite{noscale1,noscale18}. The transition to the new parameters is
described in \Sref{app1} and then, in Secs.~\ref{app2} and
\ref{app3} the particle spectrum and the SSB parameters are
derived respectively.

\subsection{\sc\small\sffamily  Half-Plane Formulation}\label{app1}

It is well-known \cite{linde} that the hyperbolic geometry is also
parameterized in the half-plane coordinates $T$ and $\bar T$ which
are related to the disc coordinates $Z$ and $\bar Z$, employed in
the main text, through the analytic transformation
\beq Z
=-\sqrt{N}\frac{T-1/2}{T+1/2}~~~\mbox{with}~~~|T|<1/2.\label{TZ}\eeq
Inserting \Eref{TZ} into \Eref{khi} for $k=0$, $\khio$ may be
expressed in terms of $T$ and $\bar T$ as follows
\beq \label{khi1} K_{\rm H0}=-N\ln\frac{T+\bar T}{(T+1/2)(\bar
T+1/2)}\,.\eeq
Upon performing a convenient \Kaa\ transformation we can show that
the model described by \eqs{khi1}{whi} is equivalent to a model
relied on the \Ka\
\beq \label{Khit} \wtilde K_{\rm H}=-N\ln\lf {T+\bar T}\rg\eeq
and the superpotential
\beq\label{whit} \wtilde W_{\rm
H}=-\sqrt{N}m(T^2-1/4)(T+1/2)^{N-2}\,.\eeq
The \Kaa\ metric, the Ricci curvature and the curvature associated
with $\wtilde K_{\rm H}$ are respectively
\beq\label{gT} \wtilde{\rm g}=\frac{N}{\lf T+\bar
T\rg^{2}},~~\wtilde R_{T\bar T}=-2\frac{\wtilde{\rm
g}}{N}~~\mbox{and}~~~\wtilde {\mathcal R}_{\rm
H}=-\frac{2}{N}\,.\eeq
Note that the last result coincides with that in \Eref{RHN}.

\subsection{\sc\small\sffamily  Hidden-Sector
Spectrum}\label{app2}

\begin{figure}[t]%
\epsfig{file=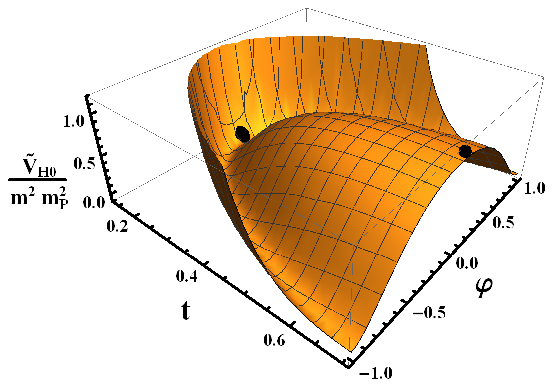,width=8.7cm,,angle=-0} \vspace*{2.3in}
\caption{\sl \small Three dimensional plot of the (dimensionless)
hidden-sector potential $\widetilde V_{\rm H0}/m^2 \mP^2$ as a
function of the parameters $t$ and $\varphi$ defined in
\Eref{Tpara}. The values $t_{\rm max}$ and $t_{\rm min}$ for
$\varphi=0$  are also depicted by thick black points.}\label{fig4}
\end{figure}

Substituting \eqs{Khit}{whit} with $N=4$ into \Eref{Vsugra} we
find the corresponding SUGRA potential which reads
\beq \label{Vht}\tVhi
=\frac{m^2}{4}\left|T+\frac12\right|^4\lf\frac{1+4|T|^2-4(T+\bar
T)}{(T+\bar T)^2}\rg^2.\eeq
To investigate further the structure of $\tVhi$, we analyze $T$ in
real and imaginary parts as follows \beq T=\lf
t+i\varphi\rg/\sqrt{2}\label{Tpara}\eeq and depict $\tVhi$ in
\Fref{fig4} as a function of these parameters for $0\leq
t\leq1/\sqrt{2}$ and $-1\leq\varphi\leq1$. We observe that
$\tVhi(\varphi=0)$ develops two extrema at $t=t_{\rm max}$ and
$t_{\rm min}$ with
\beq \label{sol} t_{\rm
max}\simeq\frac1{\sqrt{2}}~~\mbox{and}~~~t_{\rm
min}=\frac1{\sqrt{2}}\lf2-\sqrt{3}\rg,\eeq
from which $t_{\rm max}$ corresponds to a maximum whereas $t_{\rm
min}$ corresponds to a global minimum with vanishing
$\veva{\tVhi}$. Moreover, we see that the direction $\varphi=0$ is
unstable for $0<t\leq1/\sqrt{2}$ contrary to the situation in
\Fref{fig5} where the direction $\th=0$ is stabilized for all
values of $z\leq\vev{z}$.

%%%%%%%%%%%%%%%%%%%%%%%%%%%%%%%%%%%%%%%%%%%%%%%%%%%%%%%%%%%%%%%%%%%%
%%%%%%%%%%%%%%%%%%%%%%%%%%%%%%%%%%%%%%%%

If we derive the spectrum of the theory at $t_{\rm min}$ we infer
that this consists of a massless axion, a real scalar field, $\hat
t$ and $\Gr$. The masses of the two latter particles are $\what
m_t=\mzo$ and $\wtilde m_{3/2} =\mgro$ given by \eqs{mz0}{mgr0}
respectively. These masses fulfill again \Eref{strace1} where
$\wtilde m_{3/2}$ is now calculated as follows
\beq \wtilde m_{3/2}^2=4m^2\vev{\left|T+1/2\right|^4
\left|T^2-1/4\right|^2(T+\bar T)^{-4}}.\label{Tstrace}\eeq
The existence of the axion with zero mass is justified by the fact
that a $\mathbb{Z}_2$ symmetry remains unbroken. Indeed, $\tVhi$
in \Eref{Vht} is a function of $|T|^2$ and $(T+\bar T)$ and so
remains invariant under the reflection $\varphi\to -\varphi$.
Including, though, a quartic term as that emerging in the argument
of the logarithm in \Eref{khi} for $n=4$, we can generate a
non-vanishing mass $\what m_\varphi$ for $\what\varphi$. In
particular, if we employ the \Ka \beq \label{khitq} \wtilde K_{\rm
H}=-4\ln\lf {T+\bar T}+k\tm^4/4\rg~~\mbox{with}~~~\tm=T-\bar T,
\eeq we obtain \beq \what m_\varphi =3 \lf26-15\sqrt{3}\rg\sqrt{k}
m_{3/2}.\label{mtht}\eeq Moreover, alternative choices like
\beqs\bel \wtilde K_{\rm H}&=-4\ln\lf {T+\bar
T}\rg-N_k\ln\lf1+k\tm^4/N_k\rg
\\ \mbox{or}~~\wtilde K_{\rm H}&=-4\ln\lf {T+\bar T}\rg-k\tm^4\end{align}\eeqs result to a little lower mass \beq \what
m_\varphi =3\lf7-4\sqrt{3}\rg \sqrt{k}m_{3/2},\eeq without
generating any ramification either to location of the Minkowski
vacuum or to the residual particle spectrum. Apparently, our
solutions are not included in those presented in \crefs{noscale1,
noscale18}, where \Ka s of the type of $\wtilde K_{\rm H}$ in
\Eref{Khit} are considered.

\subsection{\sc\small\sffamily Soft SUSY-Breaking Parameters}\label{app3}

Another key difference of our scheme with the pure no-scale models
\cite{noscale1} is that here non-vanishing SSB parameters are
generated. This feature insists employing the present
parametrization too. To prove it, we below find the SSB parameters
involved in the potential of \Eref{vsoft}, adopting the
superpotetial in \Eref{w0} for the visible-sector fields $\phc_a$
and one of the following \Ka s
\beqs\bel
\wtilde K_1&=\wtilde K_{\rm H}+\mbox{$\sum_\al$}|\phc_\al|^2,~~~\label{tK1}\\
\wtilde K_2&=-4\ln\left(T+\bar T+k\tm^4/4-\mbox{$\sum_\al$}|\phc_\al|^2/4\right),\label{tK2}\\
\wtilde K_3&=\wtilde K_{\rm H}-N_{\rm
O}\ln\left(1-\mbox{$\sum_\al$}|\phc_\al|^2/N_{\rm
O}\right),\label{tK3} \end{align}\eeqs
taking as reference $\wtilde K_{\rm H}$ defined in \Eref{khitq}.
For low $\phc_\al$ values, the $K$'s above reduce to that shown in
\Eref{Kho}, with $\wkhi$ being identified as
\beq \label{wkt} \wkhi=\begin{cases} 1&\mbox{for}~~K=\wtilde K_1,
\wtilde K_3;\\\lf T+\bar T+k\tm^4/4\rg^{-1}&\mbox{for}~~K=\wtilde
K_2\,.\end{cases}\eeq
With the aid of \eqss{ma}{Aa}{Ba} we extract the following SSB
masses
\beqs\beq \label{mti} \wtilde m_\al=\begin{cases} \mgr&\mbox{for}~~K=\wtilde K_1, \wtilde K_3;\\
\mgr/2&\mbox{for}~~K=\wtilde K_2\end{cases}\eeq
trilinear couplings
\beq \label{Amt} A=
\begin{cases}2\lf12+7\sqrt{3}\rg\mgr&\mbox{for}~~K=\wtilde K_1, \wtilde K_3;\\
\sqrt{3}\sqrt{2+\sqrt{3}}\mgr/2&\mbox{for}~~K=\wtilde K_2\\
\end{cases}\eeq
and bilinear coupling
\beq \label{Bmt} B=\begin{cases}\lf17+10\sqrt{3}\rg\mgr &\mbox{for}~~K=\wtilde K_1, \wtilde K_3;\\
\lf\sqrt{3}+1\rg\mgr&\mbox{for}~~K=\wtilde K_2,
\end{cases}\eeq\eeqs
Comparing the above results with those in \eqs{mk1}{mk2} we remark
that $\wtilde m_\al$ are exactly the same.

As a bottom line, the SSB parameters acquire here non-vanishing
values, distinguishing further our set-up from the traditional
no-scale SUGRA \cite{noscale1, noscale13}.

%Despite these similarities, we believe that the model above
%conceptually deviates from the road of no-scale SUGRA. Indeed,
%%contrary to the traditional no-scale models of SUSY breaking , the
%complicate structure of $\wtilde W_{\rm H}$ in \Eref{whit} plays
%a crucial role in obtaining the Minkowski vacuum.

\newpage

\def\ijmp#1#2#3{{\sl Int. Jour. Mod. Phys.}
{\bf #1},~#3~(#2)}
\def\plb#1#2#3{{\sl Phys. Lett. B }{\bf #1}, #3 (#2)}
\def\prl#1#2#3{{\sl Phys. Rev. Lett.}
{\bf #1},~#3~(#2)}
\def\rmpp#1#2#3{{Rev. Mod. Phys.}
{\bf #1},~#3~(#2)}
\def\prep#1#2#3{{\sl Phys. Rep. }{\bf #1}, #3 (#2)}
\def\prd#1#2#3{{\sl Phys. Rev. D }{\bf #1}, #3 (#2)}
\def\npb#1#2#3{{\sl Nucl. Phys. }{\bf B#1}, #3 (#2)}
\def\npps#1#2#3{{Nucl. Phys. B (Proc. Sup.)}
{\bf #1},~#3~(#2)}
\def\mpl#1#2#3{{Mod. Phys. Lett.}
{\bf #1},~#3~(#2)}
\def\jetp#1#2#3{{JETP Lett. }{\bf #1}, #3 (#2)}
\def\app#1#2#3{{Acta Phys. Polon.}
{\bf #1},~#3~(#2)}
\def\ptp#1#2#3{{Prog. Theor. Phys.}
{\bf #1},~#3~(#2)}
\def\n#1#2#3{{Nature }{\bf #1},~#3~(#2)}
\def\apj#1#2#3{{Astrophys. J.}
{\bf #1},~#3~(#2)}
\def\mnras#1#2#3{{MNRAS }{\bf #1},~#3~(#2)}
\def\grg#1#2#3{{Gen. Rel. Grav.}
{\bf #1},~#3~(#2)}
\def\s#1#2#3{{Science }{\bf #1},~#3~(#2)}
\def\ibid#1#2#3{{\it ibid. }{\bf #1},~#3~(#2)}
\def\cpc#1#2#3{{Comput. Phys. Commun.}
{\bf #1},~#3~(#2)}
\def\astp#1#2#3{{Astropart. Phys.}
{\bf #1},~#3~(#2)}
\def\epjc#1#2#3{{Eur. Phys. J. C}
{\bf #1},~#3~(#2)}
\def\jhep#1#2#3{{\sl J. High Energy Phys.}
{\bf #1}, #3 (#2)}
\newcommand\jcap[3]{{\sl J.\ Cosmol.\ Astropart.\ Phys.\ }{\bf #1}, #3 (#2)}
\newcommand\njp[3]{{\sl New.\ J.\ Phys.\ }{\bf #1}, #3 (#2)}
\def\prdn#1#2#3#4{{\sl Phys. Rev. D }{\bf #1}, no. #4, #3 (#2)}
\def\jcapn#1#2#3#4{{\sl J. Cosmol. Astropart.
Phys. }{\bf #1}, no. #4, #3 (#2)}
\def\epjcn#1#2#3#4{{\sl Eur. Phys. J. C }{\bf #1}, no. #4, #3 (#2)}


\begin{thebibliography}{99}
\section*{\refname} \ignorespaces

\bibitem{nilles} H.P.~Nilles, {\sl Phys.\ Rept. } {\bf 110}, 1
(1984).

\bibitem{vafa} G.~Obied, H.~Ooguri, L.~Spodyneiko and C.~Vafa,
\arxiv{ 1806.08362}.
  %%CITATION = ARXIV:1806.08362;%%

\bibitem{lindev}  Y.~Akrami, R.~Kallosh, A.~Linde and V.~Vardanyan,
 Fortsch.\ Phys.\  {\bf 67}, no. 1-2, 1800075 (2019) [\arxiv{1808.09440}].


\bibitem{polonyi} J. Polonyi, Budapest preprint KFKI/1977/93 (1977).

\bibitem{hall} M. Claudson, L. Hall and I. Hinchliffe, {\sl Phys. Lett. B }{\bf
130}, 260 (1983).


\bibitem{rnelson} A.E. Nelson and N. Seiberg, \npb{416}{1994}{46} [\hepph{9309299}].

\bibitem{linde1} R.~Kallosh and A.~Linde, {\sl Comptes Rendus Physique }{\bf 16}, 914 (2015)
[\arxiv{1503.06785}].
%  doi:10.1016/j.crhy.2015.07.004
  %``Escher in the Sky,''


\bibitem{linde} J.J.M.~Carrasco, R.~Kallosh, A.~Linde and D.~Roest,
{\sl Phys.\ Rev.\ D }{\bf 92}, no. 4, 041301 (2015)
[\arxiv{1504.05557}].
 % doi:10.1103/PhysRevD.92.041301
%%CITATION = doi:10.1103/PhysRevD.92.041301;%%
  %``Hyperbolic geometry of cosmological attractors,''

\bibitem{su11} C.~Pallis and N.~Toumbas, \jcap{05}{2016}{no. 05, 015}
[\arxiv{1512.05657}].
%%CITATION = doi:10.1088/1475-7516/2016/05/015;%%



\bibitem{noscale} E. Cremmer, S. Ferrara, C. Kounnas and D.V. Nanopoulos,
{\sl Phys.\ Lett.\ B } {\bf 133}, 61 (1983).

\bibitem{noscale1} J.R. Ellis, C. Kounnas and D.V. Nanopoulos, {\sl Nucl. Phys. }{\bf
B241}, 406 (1984).

\bibitem{noscale18}  J.~Ellis, B.~Nagaraj, D.V.~Nanopoulos and K.A.~Olive,
\jhep{11}{2018}{110} [\arxiv{1809.10114}].

\bibitem{noscale13} J.~Ellis, D.V.~Nanopoulos and K.A.~Olive,
{\sl Phys.\ Rev.\ D }{\bf 89}, no. 4, 043502 (2014)
[\arxiv{1310.4770}].
  %``A no-scale supergravity framework for sub-Planckian physics,''


\bibitem{soft} A. Brignole, L.E. Ib\'a\~nez and C. Mu\~noz,
{\sl Adv.\ Ser.\ Direct.\ High Energy Phys. }{\bf 18}, 125 (1998)
[\hepph{9707209}].


\bibitem{masiero} G.F. Giudice and A. Masiero, {\sl Phys. Lett. B} {\bf 206}, 480 (1988).

\bibitem{raxion} J. Bagger, E. Poppitz, and L. Randall, \npb{426}{1994}{3} [\hepph{9405345}].


\bibitem{pq} R. Peccei and H. Quinn, \prl{38}{1977}{1440}.
 %%CITATION = PRLTA,38,1440;%%



\bibitem{olivegr} J.L. Evans, M. Garcia and K.A. Olive,
{\sl J. Cosmol. Astropart. Phys. }{\bf 03}, 022 (2014)
[\arxiv{1311.0052}].

\bibitem{muray} C-I. Chiang and H. Murayama, \arxiv{1808.02279}.

\bibitem{symm} G.~'t Hooft, {\sl NATO Sci.\ Ser.\ B }{\bf 59}, 135 (1980).
  %%CITATION = doi:10.1007/978-1-4684-7571-5_9;%%
  %``Naturalness, chiral symmetry, and spontaneous chiral symmetry breaking,''

\bibitem{Ibanez} See e.g. L.E.~Ib\'a\~nez and D.~L\"ust, \npb{382}{1992}{305}
[\hepth{9202046}]; D.~L\"ust, S.~Reffert and S.~Stieberger,
\npb{727}{2005}{264} [\hepth{0410074}].


\bibitem{riotto} I. Dalianis, F. Farakos, A. Kehagias, A. Riotto and
R. von Unge, {\sl J. High Energy Phys. }{\bf 01}, 043 (2015)
[\arxiv{1409.8299}].


\bibitem{dvali} G.R. Dvali, G. Lazarides and Q. Shafi, \plb{424}{1998}{259} [\hepph{9710314}].

\bibitem{univ} C.~Pallis, \jcap{04}{2014}{024}; {\bf 07}, {01(E)} {(2017)}
[\arxiv{1312.3623}]; C.~Pallis, {\sl Universe} {\bf 4}, no. 1, 13
(2018) [\arxiv{1710.05759}]; C.~Pallis and Q. Shafi,  {\sl Eur.\
Phys.\ J.\ C }{\bf 78}, no. 6, 523 (2018) [\arxiv{1803.00349}]; C.
Pallis, {\sl Eur. Phys. J. C} {\bf 78}, no. 12, 1014 (2018)
[\arxiv{1807.01154}].
  %%CITATION = doi:10.1140/epjc/s10052-018-5980-0;%%
 %%CITATION = doi:10.3390/universe4010013;%%
  %%CITATION = doi:10.1088/1475-7516/2014/04/024, 10.1088/1475-7516/2017/07/E01;%%
 %%CITATION = doi:10.1140/epjc/s10052-018-6485-6;%%


\bibitem{gaugedR} A.H. Chamseddine and H. Dreiner, {\sl Nucl. Phys. }{\bf B458}, 65 (1996)
[\hepph{9504337}]; D. J. Casta\~no, D.Z. Freedman and C. Manuel,
{\sl Nucl. Phys. }{ \bf B461}, 50 (1996) [\hepph{9507397}].


\bibitem{anomalies} G. 't Hooft, {\sl Phys. Rev. Lett. }{\bf 37}, 8 (1976).
%Symmetry breaking through Bell-Jackiw anomalies,

\bibitem{ribe} H.S.~Goh and M.~Ibe, \jhep{03}{2009}{049}
[\arxiv{0810.5773}]; Y.~Hamada \etal, \jcap{01}{2014}{024}
[\arxiv{1310.0118}].

%, K.~Kamada, T.~Kobayashi and Y.~Ookouchi

\bibitem{nilpotent} S. Ferrara, R. Kallosh and A. Linde, \jhep{10}{143}{2014}
[\arxiv{1408.4096}].
%"Cosmology with Nilpotent Superfields,"

\bibitem{astro} J.~Jaeckel and M.~Spannowsky, {\sl Phys.\ Lett.\ B }{\bf 753}, 482 (2016)
[\arxiv{1509.00476}]; J.S.~Lee, \arxiv{1808.10136}.
  %``Probing MeV to 90 GeV axion-like particles with LEP and LHC,''



\bibitem{olivepheno} J.~Ellis, M.~Garcia, D.~Nanopoulos and
K.~Olive, \jcap{10}{2015}{003} [\arxiv{1503.08867}].

\bibitem{strongpheno} E. Dudas \etal, {\sl Eur. Phys. J. C }{\bf 73}, no. 1, 2268 (2013)
[\arxiv{ 1209.0499}].
%"Strong moduli stabilization and phenomenology,"




\bibitem{kapnilles} V. Kaplunovsky and J. Louis, {\sl Nucl. Phys. }{\bf B444}, 191 (1995)
[\hepth{9502077}]; K. Choi and H.P. Nilles \jhep{04}{2007}{006}
[\hepph{0702146}].

\bibitem{gauginoHall}  L.J.~Hall and L.~Randall, {\sl Nucl.\ Phys.\ }{\bf B352}, 289 (1991).
  %%CITATION = doi:10.1016/0550-3213(91)90444-3;%%
  %``U(1)-R symmetric supersymmetry,''


\bibitem{corr} I.L. Buchbinder, S. Kuzenko and Z. Yarevskaya, {\sl Nucl. Phys. }{\bf B411},
665 (1994); M. T. Grisaru, M. Rocek and R. von Unge, {\sl Phys.
Lett. B }{\bf 383}, 415 (1996) [\hepth{9605149}].


\bibitem{buch} W. Buchm\"uller, E. Dudas, L. Heurtier and C. Wieck,
{\sl J. High Energy Phys. }{\bf 14}, 053 (2014)
[\arxiv{1407.0253}].

\bibitem{eev} E.~Dudas, T.~Gherghetta, Y.~Mambrini and K.A.~Olive, {\sl Phys.\ Rev.\ D }
{\bf 96}, no. 11, 115032 (2017) [\arxiv{1710.07341}].

\bibitem{kingpolonyi} M.C.~Rom\~ao and S.F.~King, \jhep{07}{2017}{033}
[\arxiv{1703.08333}].
  %``Starobinsky-like inflation in no-scale supergravity Wess-Zumino model with Polonyi term,''


\bibitem{ketov} Y.~Aldabergenov and S.V.~Ketov, {\sl Phys.\ Lett.\ B }{\bf 761}, 115 (2016)
[\arxiv{1607.05366}].
  %``SUSY breaking after inflation in supergravity with inflaton in a massive vector supermultiplet,''



\bibitem{polonyiproblem} G.D. Coughlan \etal, {\sl Phys. Lett. B}{\bf 131}, 59 (1983).
%"Cosmological Problems for the Polonyi Potential,"



\bibitem{adiabatic}  A.D.~Linde, {\sl Phys.\ Rev.\ D }{\bf 53}, 4129 (1996)
[\hepth{9601083}]; K.~Nakayama, F.~Takahashi and T.T.~Yanagida,
{\sl Phys.\ Rev.\ D }{\bf 84}, 123523 (2011) [\arxiv{1109.2073}].

\bibitem{queve} J.P.~Conlon, S.S.~Abdussalam, F.~Quevedo and K.~Suruliz, \jhep{01}{2007}{032}
[\hepth{0610129}].
  %``Soft SUSY Breaking Terms for Chiral Matter in IIB String Compactifications,''
  %``Vanishing Cosmological Constant in SUGRA without tuning''

\end{thebibliography}
\end{document}